\documentclass[journal, a4paper, twoside, 10pt]{IEEEtran}

\usepackage[utf8]{inputenc}
\usepackage[T1]{fontenc}

\usepackage{cite}
\usepackage{url}
\usepackage[hyperfigures,backref,bookmarks,draft=false,colorlinks=true,urlcolor=blue,citecolor=red,linkcolor=red]{hyperref}

\usepackage{graphicx}

\usepackage{amsmath}

\newcommand{\topic}{Multiple-Tree Push-based Overlay Streaming}
\newcommand{\student}{Eduardo Lidanski, Supervisor: Giang Nguyen}

\begin{document}
\title{\topic}
\author{\student}

\IEEEcompsoctitleabstractindextext{
\begin{abstract}
Multiple-Tree Overlay Streaming has attracted a great amount of attention from researchers in the past years. Multiple-tree streaming is a promising alternative to single-tree streaming in terms of node dynamics and load balancing, among others, which in turn addresses the perceived video quality by the streaming user on node dynamics or when heterogeneous nodes join the network. This article presents a comprehensive survey of the different aproaches and techniques used in this research area. In this paper we identify node-disjointness as the property most approaches aim to achieve. We also present an alternative technique which doesn't tries to achieve this but does local optimizations aiming global optimizations. Thus, we identify this property as not being absolute necessary for creating robust and heterogeneous multi-tree overlays. We identify two main design goals: robustness and support for heterogeneity, and classify existing approaches into these categories as their main focus. 
\end{abstract}}

\maketitle
\IEEEdisplaynotcompsoctitleabstractindextext

\section{Introduction}
Video streaming systems are very popular on the Internet. During video streaming, the user does not have to wait the whole video to download in order to watch it, but can watch while downloading. This allows popular events to be streamed and reach a wide audience in near real-time. Traditional video streaming is done in a client-server based architecture, in which the server maintains the whole video and users connect directly to it in order to get the stream. But this generates enormous costs for the server in terms of bandwidth requirements, which turns out to be very expensive, so a client-server approach does not scale well for a wide audience. P2P streaming applications are very popular to circumvent this problem: in P2P-based systems, each user contributes upload bandwidth by forwarding parts of the video content to other users. But this approach has some drawbacks which must be taken into account during the system design. These papers offers a comprehensive survey of actual multiple-tree overlay streaming systems, which are a special kind of P2P streaming systems, and compares them in terms of robustness and heterogeneity support. 

This paper is organized as follows. In Section \ref{sec2} necessary background is presented in order to understand the problems and challenges of tree-based streaming. Section \ref{classification} classifies the presented approaches with respect to different parameters. Sections \ref{robustness} and \ref{heterogeneity} then discuss the protocols in detail, and a thorough comparison is presented. Finally, Section \ref{conclusion} concludes the paper.

\section{Background}
\label{sec2}
As stated in the previous section, traditional server-based video streaming incurs a significant cost. Therefore, alternatives have been intensely researched in the past years. One of these alternatives is IP multicast. IP multicast is based on the Internet Protocol (IP), which has support for multicast groups. On IP multicast, routers are in charge of the packet duplication when delivering content to the multicast group.  However, it has never been widely deployed on the Internet because of the added complexity on routers and protocols. In order to widely support IP multicast, all intermediate routers on the Internet would have to be replaced. 

Because of the shown infeasibility, Application Layer Multicast (ALM) has been proposed as an alternative to IP multicast. In ALM, the multicasting functionality is shifted to the application layer instead of being handled by the IP Layer. This means that the end-nodes are in charge of the packet duplication while multicasting, not the routers as in IP multicast. The video contents are replicated and sent to other users in the end-node applications. ALM is thus a promising solution for P2P video streaming, in which end-users contribute with upload bandwidth and processing capacity to other users. In contrast to IP multicast, no changes to the intermediate routers must be done, since the end-node application layer is in charge of the content duplication.

Such ALM systems have a major drawback that must be addressed. End-nodes are, in contrast to routers, highly unreliable. Users can join and leave the system dynamically, so the ``currently active'' set of nodes is unpredictable. ALM systems must thus address this issue by maintaining and repairing the overlay when a user leaves the system or when new participants join the overlay. 

Several ALM techniques have been proposed. In the next subsection, several tree-based ALM schemes for video streaming are discussed. After that, MDC video coding is  briefly introduced.

\subsection{Tree-based topologies}
As already mentioned, ALM systems are a promising solution for P2P video distribution, in which end-users contribute with upload bandwidth and processing power to redistribute the video contents to other users. The peers in the multicast group must thus be coordinated in order to ensure that they all receive the necessary packets for correct video playback in a timely manner. The most natural way to construct such an overlay is in a tree-based scheme, in which a tree is constructed to redistribute media contents to all multicast participants. Tree-based approaches can be classified as single-tree-based or multi-tree-based systems. 

In single-tree topologies, as the name suggests, one tree is constructed, which is routed at the streaming source. Users interested in the video join the overlay tree as interior or leaf nodes. The position in which peers join the overlay is dictated by the overlay protocol. The video is then pushed from the source to the leaves of the tree. Intermediate nodes are responsible to redistributing the video content to their own children. Since the video is pushed along the tree overlay, such systems are also called push-based systems. Children contribute to the system by duplicating the packets they receive from their parents and forwarding them to their own children through unicast. 

Single-tree overlays have two big drawbacks that must be addressed for correct video redistribution. The first one is \emph{robustness}. Node dynamics affect severely the tree overlay, since node departures immediately break media delivery. All the children of the departing nodes loose significant parts of the video content until the tree is repaired, so they may have playback interruptions. \emph{Load balancing} should also be addressed. In single-tree based systems, only a small subset of peers (the internal peers) are responsible for redistributing the whole media content. 

In order to deal with these drawbacks, multiple-tree based topologies were proposed. These schemes often use the MDC video coding presented in the next section. In these schemes, one tree is created for each substream. End-users join all trees in order to receive all media. A node is placed as an internal node in \emph{only} one tree and as a leaf in all other trees. Thus, load is balanced among all participants. This is fairer than single-tree approaches where only internal nodes contribute upload bandwidth. 

\begin{figure}[bth]
\centering
\includegraphics[scale=0.5]{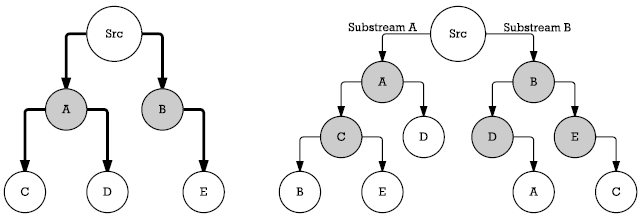}
\caption{Single-tree and multiple-tree overlays (adapted from \cite{richerzhagen:thesis})}
\label{fig:1}
\end{figure}

Because of the MDC properties, if a node fails, its children can still play the video with a lower quality. Video playback is not be interrupted. Therefore, systems using multiple-tree overlays are much more robust on node dynamics than systems using a single-tree overlay. Figure \ref{fig:1} compares a single-tree and a multiple-tree overlay. Grey nodes are responsible for forwarding video data to their children. As it can be seen in the picture, if node $A$ fails in the single-tree overlay, $C$ and $D$ will disconnect from the system, causing playback interruption. In contrast, in the multiple-tree overlay, if node $A$ fails, its children nodes $B$, $C$, $D$, and $E$ can obtain the second substream $B$, allowing them to continue playback in a lower quality without interruption. 

This paper focuses on how to construct such multiple-tree overlays and how to maintain these overlays on node dynamics. Several approaches are presented and compared.

\subsection{MDC video coding}

Multiple Description Coding (MDC) encodes media data into $n$ \emph{descriptions} or \emph{stripes} In this paper, the terms ``descriptions'', ``stripes'' or ``slices'' are used interchangeably. Any subset of these $n$ descriptions could be used to decode the video. The more descriptions available, the better the quality is of the decoded video stream with respect with the original video stream. This property of MDC can be used in streaming topologies where packet loss may happen, e.g. on node departure. The media data can still be decoded using the available descriptions without having to interrupt playback. Details of the MDC encoding may be found on \cite{Goyal01multipledescription} and \cite{Akyol0FlexibleMultiple}. In this paper, it is assumed that MDC coding is used.

\section{Classification}
\label{classification}
In this section a classification of multiple-tree based streaming systems is presented according to different parameters. 

Most approaches presented in this paper aim to achieve \emph{node-disjointness} among all multicast trees. Node-disjointness means that a node that is an interior node in a multicast tree is a leaf node in all other trees. This property ensures load balance between all participating nodes and, in combination with MDC video coding, improves resilience to node failures. 

\begin{figure}[bth]
\centering
\includegraphics[scale=0.3]{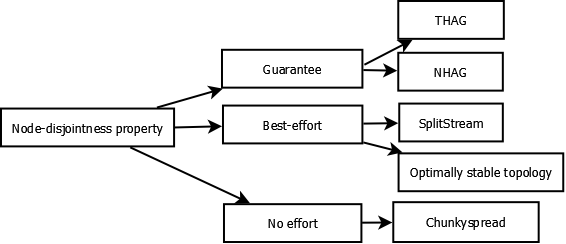}
\caption{Classification on node-disjointness}
\label{fig:node-disjointness}
\end{figure}

Figure \ref{fig:node-disjointness} shows a classification of the systems regarding node-disjointness. In the figure it is shown that some systems ensure that this property is fulfilled, while others just try to achieve it. Chunkyspread, in contrast, is the only system which does not aim to achieve it. 

\begin{figure}[bth]
\centering
\includegraphics[scale=0.3]{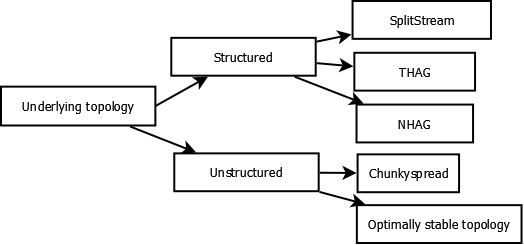}
\caption{Classification on topologies}
\label{fig:topology}
\end{figure}

The systems could also be classified regarding their underlying topology. This classification is shown on Figure \ref{fig:topology}. As shown in Section \ref{split},  SplitStream uses an existing DHT topology, while THAG and NHAG create and use their own topology. In contrast, Chunkyspread and the ``Optimally Stable Topology'' approach do not use any structured topology, but they rely on local optimization aiming global optimization.

\begin{figure}[bth]
\centering
\includegraphics[scale=0.3]{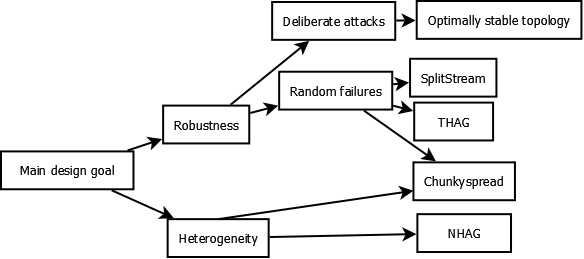}
\caption{Classification on design goals}
\label{fig:goals}
\end{figure}

Finally, the systems could be classified regarding their main design goal. The two design goals identified are \emph{robustness} and \emph{heterogeneity}. Robustness means that the system should not degrade significantly on node dynamics or deliberate attacks. Most systems focus on random node failures (node dynamics) while ``optimally stable topologies'' concentrate on deliberate attacks. Heterogeneity means that the system should support a wide range of systems, more specifically a wide range of upload/download bandwidths constraints. Figure \ref{fig:goals} show the classification. While ``Optimally stable topologies'' also supports heterogeneity, its main design goal is deliberate attack robustness, so it is shown only on this category in our classification.

\section{Approaches for achieving robustness}
\label{robustness}
\subsection{Node-disjointness}

\subsubsection{SplitStream}
\label{split}
SplitStream \cite{Castro:2003:SHM:1165389.945474} was proposed with the main goal to distribute the media content over all peers in the overlay. The system aims to construct an overlay with the aforementioned node-disjointness property.  In contrast to other systems, SplitStream does not construct its own overlay. Instead, it uses an existing DHT (Pastry \cite{Rowstron01pastry:scalable}) and a group communication system on top of this DHT (Scribe \cite{Castro02scribe:a}). To distinguish multicast groups from each other, a \emph{groupId} is associated with each multicast group. A multicast tree is then associated with each group, so each tree will have its own \emph{stripeId}.  In this section, groupIds and stripeIds are used interchangeably.

In Pastry, a message is forwarded to peers whose nodeIds share progressively larger prefixes with the message's key. These property is used in SplitStream to construct node-disjoint trees. Interior nodes share some number of digits with the tree's groupId. So for creating node-disjoint trees, groupIds are chosen such that they all start with a different digit. When dividing the video into stripes, the stripe $0$ receives stripeId $0x$, the stripe $1$ stripeId $1x$, etc. Interior nodes of stripe $0$ all have nodeIds starting with $0x$ because of the mentioned Scribe properties. This means that these interior nodes are leaves on all other trees. Thus, the node-disjointness property is achieved. 

When a node wants to \emph{join} a group, it simply routes a join message towards the groupId. This message then reaches some group member at the internal tree nodes because of Pastry's local route convergence. If the group member has not yet reached its upload bandwidth capacity limit, it accepts the joining node as a child. In contrast, if the contacted node already reached its capacity limit, it first accepts the joining node as a child, ignoring its capacity limits. Then, it re-evaluates its new set of children for selecting a child among them to reject. This re-evaluation is based on the properties of its set of children, e.g. their nodeId. The selected child is then rejected and, since the child is now an orphan, it has to find a new parent. It recursively contacts the parent's children and tries to join one of them. This method to find a parent is called \emph{push-down}. This push-down process continues down the tree until the orphan either finds a new parent or reaches the end of the tree. 

If no parent has been found during the process described, the \emph{spare capacity group} is contacted with an anycast message. The spare capacity group is a special Scribe group with a well-known groupId. It contains all nodes that have less children than their upload bandwidth capacity limit permit. If a node in the spare capacity group accepts the orphan and reaches its upload capacity, it leaves this group. An orphan is accepted only if the parent node receives the stripe the orphan is seeking for. Even after contacting the spare capacity group, it is not guaranteed that the orphan finds a parent. If that is the case, for example when the spare capacity group is empty, or where no nodes receive the stripe the orphan is seeking for, the application receives a notification that there is no forwarding capacity left in the system. 

It is important to note that, because of the spare capacity group mechanism, the node-disjointness property may be sacrificed. This happens because the new parent contacted as a result of the anycast to the spare capacity group may already be an interior node in another stripe tree. Thus, if this parent node fails, more than one stripe is lost for its children nodes. The authors show through simulation that only a small number of nodes and stripes are typically affected. They show in their simulation that the maximal number of stripes lost at a node when a single node fails (its \emph{worst case} ancestor) is $7.2$ out of $16$ total stripes. In their $40000$-node simulation the mean and median number of lost stripes when a \emph{random} node fails is always $1$. Because of this, they claim that SplitStream is very robust to random node failures.

SplitStream's robustness depends completely on the underlying Scribe and Pastry overlays, as it does not provide additional mechanisms to recover from node failures. As already mentioned, the authors show through their evaluation that SplitStream recovers quickly from node failures. But it is important to note that this recovery solely depends on the underlying overlays. 

\subsubsection{THAG}
\label{thag}
As discussed in the previous section, SplitStream aims to create multiple node-disjoint trees. Nevertheless, because of the push-down process and the anycast to the spare capacity group, it is possible that a node is a parent node in more than one tree. As a consequence, when this node fails, more than one stripe may be lost. This violates the aimed node-disjointness property. In SplitStream, node-disjointness is thus a desired property, but node-disjointness is not guaranteed. ``Topology-aware hierarchical arrangement graph'' (THAG) \cite{Tian05robustand} was proposed in order to overcome this issue. In the THAG scheme, node-disjointness is ensured. This guarantees that any host's failure in the overlay affects the data delivery in at most one multicast tree. In contrast to SplitStream, THAG provides additional mechanisms to recover from node failures (the so called ``fast switch''), and it constructs its own overlay. As already discussed, SplitStream does not, since it heavily relies on Pastry and Scribe for this purpose. 

The main idea in the THAG approach is to group all participants into a number of small Arrangement Graphs (AGs). In each AG, multiple independent multicast trees are embedded. For scalability purposes, the AGs are further organized into a hierarchical structure. The embedded trees in the AGs are, by construction, node-disjoint to each other. In the construction of the hierarchical structure, the node-disjointness property is maintained, thus ensuring node-disjointness for the whole hierarchy. 

AGs are a special kind of graphs with a set of desired properties for overlay topologies, such as strong resilience and good fault-tolerance \cite{DBLP:journals/jsa/ChenJS01}. In AGs, the address space encodes routing information in the logical addresses. A similar kind of self-routing property was already discussed in the previous section while discussing the Pastry overlay.

An AG is denoted by $A_{n,k}$, with parameters $n$ and $k$ ($1 \le k \le n$). There are $k$ symbols denoted as $X = x_1x_2\ldots x_k$, so $k$ denotes the address length. In $A_{n,2}$, nodes exist with addresses $n_{1,2}$ or $n_{2,3}$ for example. $n$ denotes the size of the AG. In THAG, $A_{n,2}$ is chosen as the structure unit to construct multicast overlays. Thus, all node addresses have length 2 and there are at most $n(n-1)$ nodes in the AG. 

AGs have important properties described in the following. An edge is constructed between nodes whose logical addresses differ by only one digit from another. So, for example, in $A_{4,2}$, node $n_{1,2}$ has connections with nodes $n_{1,3}$ and $n_{1,4}$. Trees with the node-disjointness property are directly embedded into $A_{n,2}$. In \cite{Tian05robustand}, the authors present an algorithm for embedding $n-2$ independent multicast trees into an $A_{n,2}$. In a $A_{n,2}$, a maximum of $n-2$ independent trees can be embedded. Thus, in order to construct $s$ multicast  trees for $s$ different video stripes, $n$ should be selected such that $n \ge s + 2$. If $n > s + 2$, the used trees embedded in the AG are less than the possible trees, so the unused trees can be used for further improving the system robustness.

It is important to note that, for every node in $A_{n,2}$, its parent node in each tree is also its neighbor node in the AG overlay. As a consequence, each node can determine its parent for each tree locally. Further, several AGs can be combined into a hierarchical structure preserving the node-disjointness property. In other words, a given AG can become a parent AG for other child-AGs. 

\begin{figure}[bth]
\centering
\includegraphics[scale=0.3]{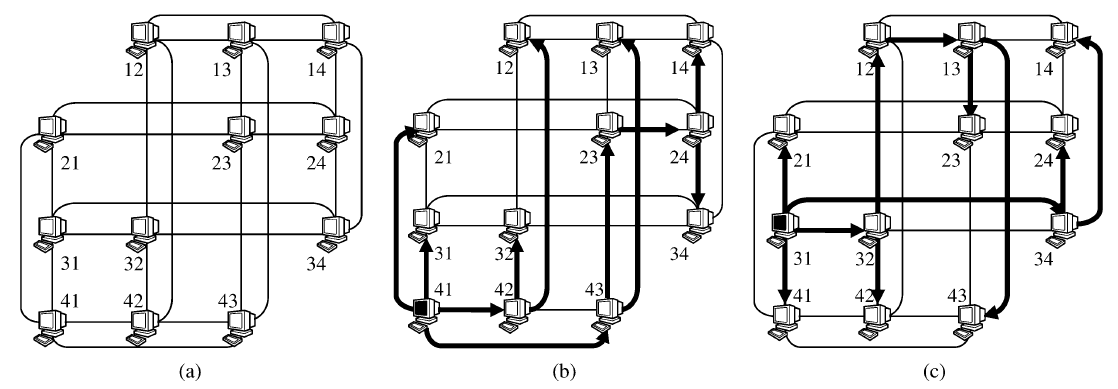}
\caption{THAG tree structure based on a) AG with $n=4$. b) multicast tree rooted at node 41. c) multicast tree rooted at node 31. (adapted from \cite{Tian05robustand})}
\label{fig:2}
\end{figure}

Figure \ref{fig:2}(a) shows an example AG $A_{4,2}$. \ref{fig:2}(b) and \ref{fig:2}(c) show the multicast trees embedded into this AG. In this figure it can be seen that the node-disjointness property is achieved. Internal nodes in one tree are leaves in the other tree.

The \emph{join} procedure works as follows. The joining node at first sends a join message to the highest AG it can enter (regarding the AG's size). Each AG has a full-leaf node that is used as an AG-entrance. The AG entrance knows the topology of the AG it is responsible for. So it knows if the AG is currently full or if there is free position for the joining node. It may also decide to replace an existing node, which then will become an orphan node in a similar way as discussed in SplitStream. If a node is rejected, either the joining node or an existing node, it will look for a new parent in the overlay. This occurs in a similar way as in SplitStream (push-down mechanism), with the rejected node communicating with the AG-entrances during the push-down mechanism. In THAG, this process is called the \emph{sinking procedure} (Locating Replacing Sinking LRS algorithm). Since AGs are organized in a hierarchical way, the rejected node will eventually locate a position in the hierarchical AGs and join the multicast streaming service. 


\subsubsection{Optimally stable streaming topologies}
\label{osst}
In contrast to the approaches already discussed, the focus in the approach proposed by \cite{Brinkmeier:2009:ODR:1550403.1550575}, Optimally Stable Streaming Topologies, is stability against deliberate denial of service (DoS) attacks. All other approaches discussed in this paper concentrate on stability on random node failures. Since the described topology is optimized against attacks, it also serves as an upper bound to the robustness of topologies, both regarding deliberate attacks and random node failures. The authors present an analytical model to examine the stability of overlay streaming topologies and describe attack strategies, e.g. perfect attacks based on global knowledge. Topologies which are optimally stable against these perfect attacks are then presented. 

These topologies base on minimalizing dependencies and keeping the relevance of nodes balanced. This avoids single nodes to become very important i.e. distributed responsibility, and so become targets of attacks. Two kind of dependencies are described: 
\begin{itemize} 
\item Amount of dependencies: the amount of other nodes that a node is dependent on should be minimalized. This minimalizes the chance of a predecessor node leaving, which in turn minimalizes the chance of quality degradation on the video because of node failures.
\item Dependency to any single node: the dependency of a node to any single other node should be minimalized in order to keep the impact of a failing predecessor as low as possible.
\end{itemize}

In order to create optimally stable streaming topologies in a distributed environment, global knowledge would be required. Global knowledge is impossible to gather in a distributed situation and under the possible message loss and node failures. Because of these reasons, the authors present a distributed procedure to create topologies \emph{close} to optimally stable topologies.

This approach is based on the following three goals:

\begin{itemize}
\item Every node should forward data in only one spanning tree.
\item The number of distinctive direct child nodes of the source and all heads should be maximized. Heads are the direct successors of the source, i.e. they are the root nodes of the multicast trees.
\item The difference of the number of children of all heads should be at most 1.
\end{itemize}

The \emph{joining} procedure is quite simple: a node joins the overlay at any node which is already part of the topology. The node can leave and rejoin the topology at any time. The result is a random topology, which is then optimized by the nodes locally. Each node analyzes its current situation and then tries to optimize it with respect to the actual costs. When joining this random node, a \emph{connect\_request} is issued for all stripes. The contacted node then starts forwarding packets for all stripes. Using local optimizations, specifically the $K_1$ function introduced later, the overlay is optimized locally so that after optimization the node only forwards its preferred stripe.

For all outgoing links of a node, four different cost metrics $K_1$, $K_2$, $K_3$, and $K_4$ are defined. These are combined to a total cost function $K$. $K$ is calculated by every forwarding node for evaluation and optimization of the stability of its actual local situation. This local optimization at all nodes aims at the optimization of the whole topology.  

Each node $v$ calculates the total cost of the edges $e=(v,w)$ to its children in the spanning tree of stripe $i$: $K(v,w,i) = \sum\nolimits_{j=1}^4 s_j \cdotp K_j(e,i)$, where $s_j$ is the weight factor of the cost function $K_j$. The cost functions $K_1$, $K_2$, $K_3$, and $K_4$ are defined next.

A node needs to choose the stripe it wants to forward. This stripe preference should not be static. Thus, every node selects the stripe which it forwards to the most children as its preferred stripe, and assigns higher costs to all outgoing edges in the spanning trees of different stripes.
\[
K_1(v,i) := 1 - \frac{\text{fanout}_{T_i}(v)}{c^+(v)}
\]
where \emph{fanout}$_{T_i}(v)$ is the number of outgoing edges of node $v$ in the spanning tree $T_i$ and $c^+(v)$ is the upload bandwidth capacity available. $c^+(v)$ is calculated in terms of the video bitrate. So if the video bitrate is e.g. $3kbit/s$, and if the node $v$ can forward 3 times this bitrate, then $c^+(v)=9kbit/s$.  
Thus, $K_1$ will be low for links forwarding the preferred stripe and high for all other links, i.e. if a node forwards stripe $k$ many times already (high fanout), then it the link to this stripe will get a low cost. Other stripes get a high cost. Through optimization of $K_1$ only one stripe is forwarded by all nodes in the overlay in the best case.

In each spanning tree, only a subset of the participants has to forward the data. All other participants receive the data, but don't forward it. So, for each spanning tree, the forwarding nodes should be placed near the source, and other nodes as far as possible, as leaves. This keeps spanning trees low and also permits easy location of the available bandwidth. Thus, edges to nodes which prefer to forward data in the considered stripe are assigned low cost $K_2$, while edges to all other nodes a high cost. 

\[
K_2(v,w,i) := \left\{
        \begin{array}{ll}
            0 & \quad w \text{ can forward in stripe i,} \\
            1 & \quad \text{else.}
        \end{array}
    \right. 
\]

A node has to level the number of successors of its child nodes. Child nodes $w$ with a high number of children should not be put deeper in the tree. This avoids that the average depth of the nodes in the trees is high, which avoids higher dependencies. This improves robustness in case of a node failure, as mentioned earlier, and also avoids that an attacker identifies nodes with high dependencies in order to cause high damage to the system. So balancing the topology is achieved through

\[
K_3(v,w,i) := \frac{\left(\frac{\text{succ}_{T_i}(v)}{\text{fanout}_{T_i}(v)} - 1 \right) - \text{succ}_{T_i}(w)}{\left(\frac{\text{succ}_{T_i}(v)}{\text{fanout}_{T_i}(v)}-1\right)}
\]

$K_3$ will thus be high for outgoing links to nodes with many successors (also indirect successors), and low for outgoing links to nodes with a low number of successors. In the distributed environment, the number of successors succ$T_i$ of a node is lazily gathered as a reverse multicast.

In some cases, a node has to forward data in more than one spanning tree. In such cases, it is important that the direct dependency to each child is minimized. $K_4$ thus evaluates the number of direct connections between a node $v$ and each node $w$ of its children and tries to minimize it:

\[
K_4(v,w) := \frac{\text{fanout}_T(v,w)}{k}
\]

where $\text{fanout}_T(v,w)$ is the number of stripes in which $v$ directly forwards the stripe to $w$ and $k$ is the total number of stripes.

As mentioned earlier, the four presented cost functions are combined to a total cost function $K$. Optimization of the total cost function on each node locally leads to optimization of the stability of the complete topology. The optimization is done thus locally, by rearranging direct links. Nodes only optimize with respect to their children nodes. An distributed algorithm to create these optimizations is also presented in \cite{Brinkmeier:2009:ODR:1550403.1550575}.

For evaluating the approach, the authors compare the stability of a optimally stable topology with both the proposed procedure to create topologies close to optimally stable topologies and existing ALM approaches under attacks. An optimal attack based on global knowledge is used. The existing ALM topology used is a DAG topology with BCBS \cite{strufe06bcbs}. The video stream is partitioned in $k=4$ different stripes. The total disruption of the service is reached after attacking a subset of 8\% of all nodes in both the optimally stable topology as on the stable topology generated using the cost-based approach. Its stability almost matches the stability of the optimally stable topology. They both are much more stable than the DAG topology, where the total disruption of the service is reached after attacking a subset of 2.5\% of the nodes. So their robustness in respect to optimal attacks is much greater than the DAG topology. 

Compared to SplitStream, the authors conclude that, while SplitStream is stable to random node failures, it cannot be stable toward optimal attacks. This is due to the fact that SplitStream creates unbalanced and rather high topologies.

The authors further show that in their proposed system, in average, 6\% of all nodes are internal nodes in more than one spanning tree, thus forwarding more than one stripe. An overall minimum of 4\% and an overall maximum of 10\% of all nodes forward more than one stripe. Thus, the authors conclude, the node-disjointness property can be relaxed, since the violation of this property does not cause the topologies to be less stable. This conclusion is also observed in the system presented in the next section, in which the Chunkyspread approach does not even try to reach the node-disjointness property. 

\subsection{Chunkyspread}
\label{chunkyspread}
Chunkyspread \cite{Venkataraman:2006:CHU:1317535.1318351} \cite{Venkataraman06chunkyspread:multi-tree} is an unstructured, multi-tree based multicast system. 
In contrast to the systems presented previously, Chunkyspread does not impose any structure by design (thus \emph{unstructured}). Each peer joins the system in a \emph{random} position with local optimizations afterwards. As shown in this section, after exchanging neighbor information, each node joins all necessary slides, so different streaming trees are created. Even though Chunkyspread is tree-based, the system is unstructured because of the random graph joining operation. 

Unlike the other approaches presented in this paper, Chunkyspread doesn't try to achieve node-disjointness. This property, as also explained in Section \ref{osst}, is not necessary for creating robust multi-tree systems, as it is hard to achieve even in systems like SplitStream, especially in heterogeneous environments. This is due to the fact that, in heterogeneous environments, the pushdown operations and the anycast to the spare capacity group occur more frequently, which leads to repeated node disconnections. In SplitStream, because of the underlying DHT topologies, ID-based constraints are preferred over load-based constraints. Refer to Section \ref{split} for more details.

To \emph{join} a multicast group, nodes must first contact a rendezvous node at a well-known location, like in other P2P systems. This rendezvous node must know at least one participant of the multicast group. Afterwards, the joining node participates in a continuously running distributed algorithm called Swaplinks \cite{conf/infocom/VishnumurthyF06}, which enables random nodes discovery over the existing graph. So the joining node then joins the random node discovered using Swaplinks.  Swaplinks enables fine-grained statistical control over the node degree of the nodes, as well as the probability in which one node will be (randomly) selected. The idea here is that nodes with higher bandwidth capacity should have more neighbors to transmit descriptions, and nodes with lower load should have proportionally fewer neighbors. During the joining process, bloom filters \cite{WHITAKER02forwardingwithout} are used to avoid loops.

Each node periodically advertises to all of its neighbors information, so they can decide which nodes are appropriate parents for each slide. After this calculation, they can join these nodes, thus joining the necessary slice streams. The calculation is based in the parameters explained next. 

The join process takes the following parameters, among others:
\begin{itemize}
\item \emph{Target Load TL}:  volume that the participant would prefer sending at steady state. The expectation is that the volume sent during the steady state is near the target load. 
\item \emph{Maximum Load ML}: absolute maximum volume that the participant will transmit at any time.
\end{itemize}

\begin{figure}[bth]
\centering
\includegraphics[scale=0.5]{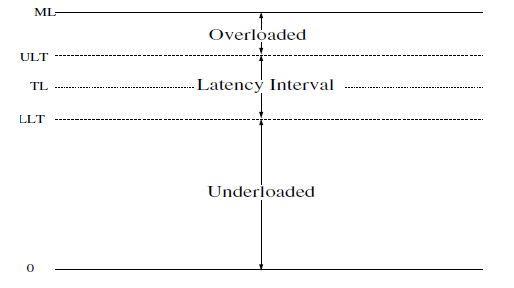}
\caption{Chunkyspread latency thresholds (adapted from \cite{Venkataraman:2006:CHU:1317535.1318351})}
\label{fig:3}
\end{figure}

Based on these parameters, and using other system-wide parameters, the following values are calculated:
\begin{itemize}
\item \emph{Lower Latency Threshold LLT}: lower edge of the latency range.
\item \emph{Upper Latency Threshold ULT}: upper edge of the latency range.
\end{itemize}
These values aim at balancing load with regard to each node's given target load. Figure \ref{fig:3} shows these latency thresholds. The load balancing mechanism works as follows. As long as the participant's load is outside the LLT-ULT range, it is either underloaded, or overloaded, respectively, so the system adjusts to move the load within this range. If a node $X$ is underloaded, other nodes will try to become a child of $X$. This increases $X$'s load. If $X$ is overloaded, its existing children will try to find other parents, which reduces $X$'s load. Once node's loads are within the LLT-ULT range, the system no longer tries to balance load, since the node is ``satisfied'', but rather tries to optimize latency. Whenever a change of parents increases latency by a certain threshold without causing the load to leave the LLT-ULT range, this change is done. So, similarly to the approach on ``Optimally stable streaming topologies'' presented in the previous section, a random graph is first created and then locally optimized. 

In order to evaluate Chunkyspread's robustness against node failures, the authors use the ``recovery duration'' metric, which is defined as the time duration between the instant when nodes \emph{detect} failures of their neighbors until they get \emph{reconnected} to the trees. The recovery duration of descendants of the disconnected node, which are also disconnected from the tree, also get accounted, since an ancestor is trying to recover on their behalf. The slice number in the simulations is $16$. 

First, robustness is simulated on \emph{bursty failures}, in which 10\% of the 5000 nodes fail at the 30th instant, at various levels of slice redundancy. Slice redundancy means streaming some slice copies, thus redundant, along with the irredundant slices. The simulation shows that both protocols (SplitStream and Chunkyspread) recover quite fast, with 90\% recovery times of about 5s in Chunkyspread and 8s in SplitStream. On adding redundant slides, there is a drastic improvement in the recovery times. With a redundancy of 3 slices, more than 50\% of the Chunkyspread nodes don't get disconnected at all, and the recovery duration has a maximum of 2.5s. 

When 50\% of the nodes fail at the same instant, Chunkyspread performs also better than SplitStream. With Chunkyspread, the 90\% recovery time is 10s, while it is at least 15s for SplitStream. With redundancy, the recovery duration is again improved: the 95\%  recovery time is just 5s for Chunkyspread with 3 redundant slides. The authors conclude that, since in SplitStream long paths are created, this affects its robustness to node failures. 

The authors also evaluate Chunkyspread on more realistic scenarios, that is, a churn scenario. They modify the buffer sizes of the respective nodes. With no buffer at all, the 90\% recovery time value is 20s, and this value decreases as buffer size is increased. For example, with a 5 second buffer size, 85\% of the slices are not disrupted at all, and the 90\% disruption duration is 1 second. So most of the disruptions are of short duration and can be recovered using buffers of small sizes.

\section{Approaches for supporting heterogeneity}
\label{heterogeneity}
In the systems presented in the previous section, robustness is the main focus. Now two systems will be discussed which support heterogeneity well: NHAG and Chunkyspread. Chunkyspread was already presented in Section \ref{chunkyspread}. As it also has good heterogeneity support, it is reviewed briefly here and its evaluation regarding heterogeneity is presented. 
\subsection{NHAG}
In Section \ref{thag} THAG was presented and discussed. A big problem of THAG regarding heterogeneity is that a constant AG size is used, which creates difficulty in delivering descriptions correctly in heterogeneous networks. In \cite{Kobayashi:2009:RES:1652959.1652973} Network-aware Hierarchical Arrangement Graph (NHAG) is presented. Its main purpose is to enhance THAG to change the AG size dynamically, thus supporting heterogeneous networks. Since NHAG bases on THAG, node-disjointness is also ensured.

In THAG, the required node upload bandwidth needed for transmission is mainly determined by the AG size. So if a node is trying to join a tree, but its actual upload bandwidth is less than the required upload bandwidth (determined by the AG size), it will not be able to forward all descriptions to its children. A result of this is that they will get a lower video quality, depending on how many descriptions are not being able to be forwarded. The main problem, as mentioned before, is that in THAG the AG size is constant. NHAG solves this problem allowing nodes join an existing AG with the appropriate size regarding their upload bandwidth.   

\begin{figure}[bth]
\centering
\includegraphics[scale=0.5]{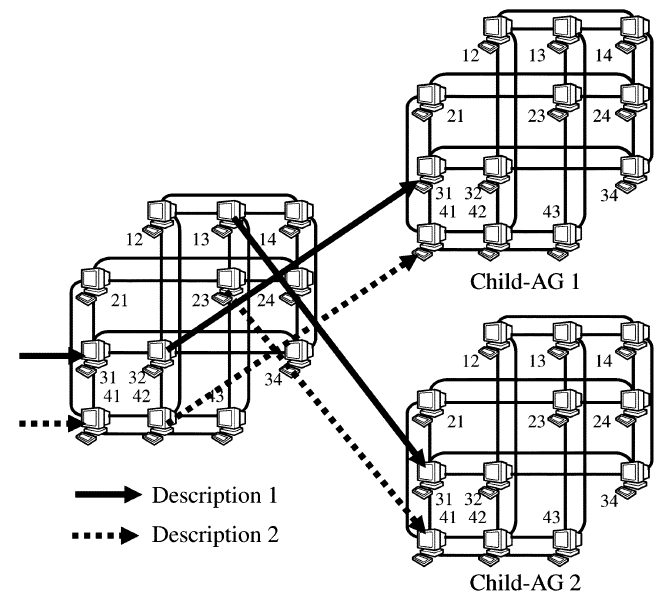}
\caption{Hierarchical AG in THAG (adapted from \cite{Kobayashi:2009:RES:1652959.1652973})}
\label{fig:4}
\end{figure}

As in THAG, an AG $A_{s,2}$ is used as the basic overlay unit. $s$ denotes the AG size. An AG with size $s$ permits $s(s-1)$ nodes to participate. If the AG is full, new nodes join to child-AGs in a hierarchical way. Figure \ref{fig:4} shows this AG hierarchy. As the figure shows, $s$ has a constant size in all AGs ($s=4$). Thus, the same number of descriptions must be forwarded by all forwarding nodes. The number of descriptions a forwarding node has to forward in an AG with size $s$ is $s-2$. In the example, if a node joins which has only upload capacity for 1 description, the other description will be discarded, and its children will get video with less quality, although they can maybe forward both descriptions.

\begin{figure}[bth]
\centering
\includegraphics[scale=0.5]{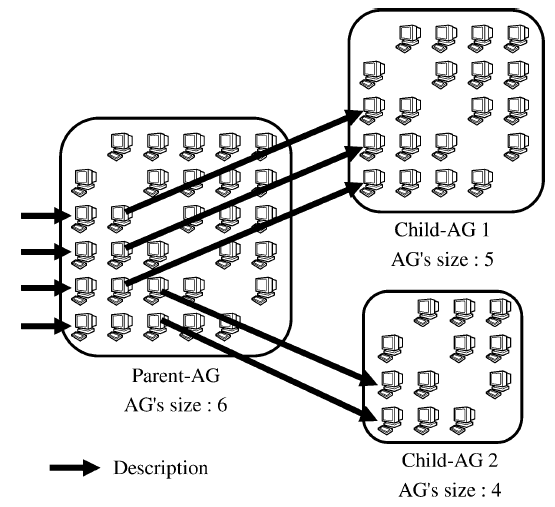}
\caption{Hierarchical AG in NHAG (adapted from \cite{Kobayashi:2009:RES:1652959.1652973})}
\label{fig:5}
\end{figure}

In NHAG, each nodes calculates the requested AG size based on its available upload bandwidth, and searches for an appropriate AG based on this size when \emph{joining} the overlay, which contrasts to THAG, where size is fixed.The AG entrance determines if the joining node is available to join its AG based on its requested size. If the AG entrance decides the node is not able to join, it forwards it to its child AG with the nearest requested size. If there is no appropriate child AG, and the AG is permitted to create a child AG, the child AG is created and the joining node enters. So the joining procedure is basically a \emph{push-down} procedure, until the joining node finds an appropriate AG or an appropriate AG is created at the bottom of the tree. Figure \ref{fig:5} shows an example of a hierarchical NHAG structure. The parent size is 6, so 4 descriptions are being forwarded. Only nodes with appropriate upload bandwidth are allowed to join this AG. Nodes with smaller bandwidths are only allowed to join its child-AGs. The main difference with THAG is that in NHAG the AG size is flexible, while in THAG the size is fixed. This is of course taken into account during the joining NHAG procedure.

In order to evaluate NHAG, the metric ``bandwidth satisfaction rate'' is studied, among others. It is defined as the ratio between the requested streaming rate and the received streaming rate. If all nodes have the same bandwidth, THAG performs equally well as NHAG, because this is the ideal case. Both are almost 1, and are above the performance of SplitStream, which is 0.8. Using different bandwidths, as expected, NHAG performs significantly better than THAG and SplitStream. NHAG continues to be almost 1, close to the ideal case. But THAG has a big performance degradation: streams received by each node were less than half of the required quality. So this indicates that NHAG delivers the required number of descriptions, not as THAG (between 0.3 and 0.6) or SplitStream, which is still 0.8. Similar results were deduced from the other metrics defined.

\subsection{Chunkyspread}
\label{chunkyspread2}
Chunkyspread is presented in Section \ref{chunkyspread}. Chunkyspread's focus is not only to be a robust system against (random) node failures, but also to widely support heterogeneity. As already mentioned, each node has a target load (TL) parameter when joining the overlay, which allows the system to calculate its LLT-ULT window. This ensures that heterogeneous nodes can join the overlay and that their target load is respected. Local optimizations take place to balance load according to each node's TL. If a node is overloaded or underloaded, neighbors are exchanged in order to correct this.

Simulation was done with moderate level of heterogeneity, representing e.g. users behind dial-up modems until broadband. In order to evaluate Chunkyspread's heterogeneity support, the metric ``Excess Load Percentage'' is further introduced. It quantifies to which extend nodes reach their target load. A value of 0\% indicates that the node has perfectly reached its TL, while a value of -100\% indicates that the node has zero load. The maximum value is 50\% in the simulations. Two different scenarios are compared: the joining scenario and the static scenario, where a steady state is reached. In \emph{Lat0} performance is quite good: more than 80\% of the nodes reach exactly their TL in the static scenario while around 90\% of the nodes reach their TL in the join scenario. In \emph{Lat2} almost 90\% of the nodes are within 25\% of their TL values in both the join and static scenarios, so they also perform well. So Chunkyspread performs well regarding heterogeneity, which is shown in the simulations.

\section{Conclusion}
\label{conclusion}
In this paper, a comprehensive and in-depth survey on Multiple-Tree Push-based Overlay Streaming was given from an algorithmic point-of-view. Different approaches were presented and their performance and methods were compared regarding their robustness and heterogeneity support. Common characteristics of the methods were identified and pointed out. Their evaluation metrics and results were thoroughly discussed and compared. 

Node-disjointness was identified as the property most of the presented systems aim to achieve. Some of them ensure this property, some of them only try to achieve it without giving any guarantees. In contrast to the node-disjointness approach, Chunkystream was presented, which does not try to achieve it, showing that this property is not absolutely necessary to have a robust and heterogeneous-supporting system.

\bibliographystyle{plain}
\bibliography{p2p}

\end{document}